\documentstyle[12pt,aasms4]{article}
\def\gray{{$\gamma$-ray\ }}
\def\grays{{$\gamma$-rays\ }}
\def\etal{{\it et al.\ }}
\begin{document}

\title{PREDICTED EXTRAGALACTIC TeV GAMMA-RAY SOURCES}
\author{F. W. Stecker}
\affil{Laboratory for High Energy Astrophysics, NASA Goddard Space 
Flight Center, Greenbelt, MD 20771, U.S.A.}
\authoremail{stecker@lheavx.gsfc.nasa.gov}

\author{O.C. De Jager}
\affil{Physics Department, Potchefstroom University, Potchefstroom 2520,
South Africa}
\authoremail{okkie@fskocdj.puk.ac.za}

\author{M. H. Salamon}
\affil{Physics Department, University of Utah, Salt Lake City, UT 84112, U.S.A.}
\authoremail{salamon@cosmic.physics.utah.edu}

\begin{abstract}

We suggest that low-redshift XBLs (X-ray selected BL Lacertae objects)
may be the only extragalactic \gray sources observable at TeV
energies. We use simple physical considerations involving synchrotron 
and Compton component spectra for blazars to suggest why the observed TeV 
sources are XBLs, whereas mostly RBLs and FSRQs are seen at GeV energies. 
These considerations indicate that the
differences between XBLs and RBLs cannot be explained purely as 
relativistic jet orientation effects.
We note that the only extragalactic TeV sources which have been observed are
XBLs and that a nearby RBL with a very hard spectrum in the GeV range has not
been seen at TeV energies. We also note that of the 14 BL Lacs observed
by EGRET, 12 are RBLs, whereas only 2 are XBLs.
We give a list of nearby XBLs which we consider to 
be good candidate TeV sources and predict estimated TeV fluxes for these
objects.

\end{abstract}


\section{Introduction}

Over 50 blazars have been detected as \gray sources in
the GeV energy range by the {\it EGRET} detector on the {\it Compton Gamma-Ray
Observatory} (Fichtel, \etal 1994; Thompson, \etal 1995, 1996). 
In contrast,
only two or three blazars have been detected at TeV energies, only one of
which is a detected GeV source. There are many {\it EGRET} blazars with 
differential photon spectra which are $E^{-2}$ power-laws or flatter. These
sources would be detectable by telescopes such as the Whipple telescope
in the TeV energy range, assuming that their spectra extrapolate to TeV
energies. In this paper, we address the questions: (1) Why has only one of the
{\it EGRET} sources has been detected at TeV energies?, and (2) Which blazars
are likely to be TeV sources?

We have already addressed part of this problem by pointing out
the critical effect of absorption of high energy \grays between the source
and the Earth by pair-production interactions with the intergalactic infrared
background (Stecker, De Jager \& Salamon 1992)
In a series of papers (Stecker \& De Jager 1997 and references therein), 
we have shown that \grays with
energies greater than $\sim 1$ TeV will be removed from the spectra of sources
with redshifts $>$ 0.1. 
Absorption
effectively eliminates flat spectrum radio quasars (FSRQs) as TeV sources.
The nearest {\it EGRET}
quasar, 3C273, lies at a redshift of 0.16. This source is also a
``mini-blazar'' which, in any case, has a steep spectrum at GeV energies.
The next closest {\it EGRET} quasar, 1510-089, has a redshift of 0.361.
At this redshift, we estimate that more than $\sim$ 99\% of the original flux 
from the source will be absorbed at TeV energies (Stecker \& De Jager 1997). 
Although the source spectra of FSRQs may not extend to TeV energies,
their distance alone makes them unlikely candidates as TeV sources. Therefore,
we consider here the more nearby blazars, which are all BL Lacerate
objects.

\section{Synchrotron and Compton Spectra of XBLs and RBLs}

An extensive exposition of blazar spectra has recently been given by
Sambruna, Maraschi \& Urry (1996). The spectral energy distributions
(SEDs) of blazars were considered by type. With the sequence FSRQs,
RBLs, XBLs, they found a decreasing bolometric luminosity in the radio to
X-ray region and an increasing frequency for the peak in the SED of the
source. Two alternative explanations have been proposed the explain this.
There is the ``orientation hypothesis'', which states that these sources
(or at least the BL Lacs) have no significant physical differences 
between them; rather the differences in luminosity and spectra result from 
relativistic beaming effects, with XBLs jets being observed with larger
angles to the line-of-sight than RBLs (Maraschi, \etal 1986; 
Ghisellini \& Maraschi 1989; Urry, Padovani \& Stickel 1991; 
Celotti, \etal 1993). In the alternative interpretation, the differences
between RBLs and XBLs must be attributed, at least in part, to real physical
differences (Giommi \& Padovani 1994; Padovani and Giommi 1995; Kollgaard,
Gabuzda \& Feigelson 1996; Sambruna, \etal 1996).

To understand the spectra of blazars, their SEDs are 
broken into two parts. The lower
frequency part, which can be roughly described by a convex parabolic
$\nu F_{\nu}$ spectrum, is generally considered to be produced by synchrotron
radiation of relativistic electrons in the jet. The higher energy part, which
includes the \gray spectrum, is usually considered to 
be produced by Compton radiation
from these same electrons. In the SEDs of XBLs, the X-ray emission comes from
the high energy end of the synchrotron emission, whereas in RBLs the X-ray
emission is from Compton scattering. This situation produces a bimodal
distribution in the broad-range radio to X-ray spectral index, $\alpha_{rx}$,
which can be used to classify BL Lac objects as XBL-like or RBL-like, or
alternatively HBLs (high frequency peaked BL Lacs) and LBLs (low frequency
peaked BL Lacs) (Padovani \& Giommi 1995, 1996; Sambruna, \etal 1996;
Lamer, Brunner \& Staubert 1996).

If real differences exist between RBLs and XBLs, one might suspect that XBLs
are more likely to be TeV sources than RBLs.
This is because in XBLs (HBLs), there is evidence from the synchrotron 
SEDs that relativistic electrons are accelerated to higher energies than in
RBLs (LBLs) ({\it e.g.}, Sambruna, \etal 1996). These electrons, 
in turn, should Compton scatter to produce higher energy \grays in XBLs 
than in RBLs. 

In fact, of the over 50 blazars seen by {\it EGRET} in the GeV range, 
including 14 BL Lacs (based on the observations given by Thompson, 
\etal 1995, 1996; Vestrand, Stacy \& Sreekumar 1995 and Fichtel, \etal 1996),
only two, {\it viz.} Mrk 421 and PKS 2155-304, are 
XBLs.\footnote{It is not clear whether the physics of the sources favors
RBLs as GeV sources or whether this is a demographic effect. Observed 
RBLs may be an order of magnitude more abundant than XBLs (Padovani \&
Giommi 1995); however this may be due to selection effects 
(Urry \& Padovani 1995; see also Maraschi, Ghisellini, Tanzi \& Treves 1986).} 
In contrast, {\it only} XBLs have been seen at TeV energies.  
Thus, the \gray observations lend further support to the LBL-HBL 
spectral difference hypothesis. We will consider this point quantitatively 
below.

\section{BL Lacertae Objects as TeV Gamma-Ray Sources}

In accord with our estimates of intergalactic absorption,
the only extragalactic TeV
\gray sources which have been reported are nearby BL Lac objects. 
The GeV \gray source Mrk 421, 
whose redshift is 0.031, was the first blazar detected at TeV energies
(Punch, \etal 1992). A similar BL Lac object, Mrk 501, whose redshift is
0.034, was detected more
recently (Quinn, \etal 1996), although it was too weak at GeV energies to
be detected by {\it EGRET}. Another BL Lac object, 1ES2344+514, whose 
redshift is 0.044, was recently
reported by the Whipple group as a tentative detection (Schubnell 1996). This
could be the third BL Lac object at a redshift less than 0.05 detected at
TeV energies. 

These observations are suggestive when considered in the
context of radio and X-ray observations of BL Lac objects.
If $\log (F_{X}/F_{r}) < -5.5$ for a BL Lac object, 
the source falls in the observational category of a radio-selected
BL Lac object (RBL), whereas if $\log (F_{X}/F_{r}) > -5.5$, the object
is classified as an X-ray selected BL Lac (XBL) (Giommi \& Padovanni
1994). Using this criterion,
{\it only XBLs have been detected
at TeV energies, whereas the RBL ON231 (z=0.1), with the hardest observed GeV 
spectrum (Sreekumar, \etal 1996), was not seen at TeV energies.}
We will show below that this result may be easily
understood in the context of simple SSC models. We further predict that
only nearby XBLs will be extragalactic TeV sources.

\section{SSC Models of BL Lacs}

The most popular mechanisms proposed
for explaining blazar \gray emission have involved either (1) 
the SSC mechanism, {\it viz.}, Compton
scattering of synchrotron radiation in the jet with the same electrons 
producing both radiation components (Bloom \& Marscher 1993 and references
therein), or (2) Compton scattering
from soft photons produced external to the jet in a hot accretion disk around
a black hole at the AGN core (Dermer \& Schlickheiser 1993), possibly
scattered into the jet by surrounding clouds (Sikora, Begelman
\& Rees 1994).

During the simultaneous X-ray and TeV flaring of the XBL Mrk421 in May
of 1994, it was observed
that the flare/quiescent flux ratios were similar for both X-rays and
TeV $\gamma$-rays, whereas the flux at and below UV frequencies 
and that at GeV energies remained constant. This observation can be understood in the context of an SSC model with the high energy tail of the radiating 
electrons being enhanced during the flare and the low energy electron spectrum
remaining roughly constant (Macomb, \etal 1995, Takahashi, \etal 1996). 
It is plausible to assume that the SSC mechanism operates generally in BL
Lac objects, since these objects (by definition) usually do not show evidence
of emission-line clouds to scatter external seed photons.


The fact that the TeV photons did not flare much more dramatically than the
X-rays implies that the enhanced high-energy electrons were scattering off
a part of the synchrotron SED which remained constant (Takahashi, \etal 1996).
This leads to the important conclusion that 
the TeV $\gamma$-rays are not the result of the inverse Compton scattering off
the X-rays, even though the synchrotron-produced luminosity peaked in the
X-ray range. 

This observation can be understood if the TeV
$\gamma$-rays were produced by Compton scattering off
photons in the UV and optical regions of the SED in which the luminosity
remained constant during the flare. This situation could have occurred during 
the flare if scatterings off optical and UV photons
occurred in the Thomson regime whereas scatterings off the more 
dominant X-rays would have been suppressed by being in the 
Klein-Nishina (KN) range.
We therefore deduce that during the flare the transition between
the Thomson and KN regimes occurred at a soft photon 
energy of $\sim$ 10 eV. 
Thus, scatterings off X-ray photons would have occurred in the extreme
KN limit. 

The boundary between Compton scattering in the Thomson and KN
limits is given by the condition 
$\epsilon E_{\gamma}/\delta ^2m^2c^4 \sim 1$, where $\epsilon$ is the energy
of the soft photon being upscattered and $E_{\gamma}$ is the energy of the
high-energy \gray produced and $\delta = [\Gamma(1-\beta\cos\theta)]^{-1}$
is the Doppler factor of the blazar jet. 
(We denote quantities in the rest system of the
source with a prime. Unprimed quantities refer to the observer's frame.) 
The factor of $\delta ^2$ results from the 
Doppler boosting of both photons from the rest frame of the emitting 
region in the jet. 
According to the above condition, the Doppler factor which 
produces a Thomson-KN transition
for soft photons near 10 eV is given by
\begin{equation}
\label{doppler}
\delta \approx 6\epsilon_{10}^{1/2}E_{\rm TeV}^{1/2}
\end{equation}
where $\epsilon_{10}=(\epsilon/10\: {\rm eV})$  and $E_{\rm TeV}=
(E_{\gamma}/1\: {\rm TeV}) $.\footnote{This value of $\delta$ is 
consistent with the condition that the jet be transparent to \grays
(see, {\it e.g.}, Mattox, \etal 1996).}

From this condition, it follows that the Lorentz factor of the scattering
electron in the source frame $\gamma_e'$, and the magnetic field
strength $B^{\prime}$, obtained from
the expression for the characteristic synchrotron frequency
$\nu_{\rm s}' \simeq 0.19 (eB'/m_{e}c)\gamma_{e}'^2$ of the soft photon,
are given by
\begin{equation}
\label{gamma}
\gamma_{e}'\simeq 3\times 10^5\epsilon_{10}^{-1/2}E_{\rm TeV}^{1/2}
\qquad\mbox{from}\qquad E_{\gamma}\sim\frac{4}{3}\gamma_{e}'^{2}\epsilon
\end{equation}
and
\begin{equation}
\label{B}
B^{\prime}
\simeq 0.2\epsilon_{\rm keV}\epsilon_{10}^{1/2}E_{\rm TeV}^{-3/2}\;\;{\rm G}
\end{equation}
where $\epsilon_x=1\epsilon_{\rm keV}$ keV is the characteristic 
X-ray synchrotron photon energy $h\nu_{s}$, resulting 
from electrons with energy
$\gamma_{e}'mc^2$ in a B-field of strength $B^{\prime}$. 
Taking $\epsilon_{10}$, $\epsilon_{\rm keV}$ and $E_{\rm TeV}$ 
equal to unity in 
eq.(\ref{B}),
we obtain a value of
$B^{\prime}\sim 0.2$ G, which is 
consistent with other estimates (Takahashi, \etal 1996).

For Mrk 421 we find that the ratio of bolometric Compton to synchrotron
luminosities $L_{\rm C}/L_{\rm syn}=U_o'/U_B'\sim 1$, 
where $U_o'$ is the rest
frame energy density in the IR to UV range (that of the seed photons),
and $U_B'=B'^2/8\pi$ is the magnetic energy density.
From this analysis we can also obtain an estimate for the 
size of the optical emitting region, $r^{\prime}$, by noting that
\begin{equation}
\label{uo}
U_o'=\delta^{-4}L_o/4\pi r'^2c 
\end{equation}
({\it e,g}, Pearson \& Zensus 1987), where $L_{o}$ is 
the luminosity
of the source in the optical-UV range $\sim 2\times 10^{44}$ erg s$^{-1}$.
From this, one obtains

\begin{equation}
\label{r'}
r' \sim 2\times 10^{16}
\epsilon_{10}^{-3/2}E_{\rm TeV}^{1/2}
\epsilon_{\rm keV}^{-1}\;\;{\rm cm},
\end{equation}

The optical variability 
timescale, given by $\tau_o\sim r'/c\delta$,
is much longer than the X-ray and TeV flare timescales. This implies that
during the flare, impulsive acceleration of the high-energy tail of the 
relativistic electron
distribution occurred over a much smaller region than that occupied by the
bulk of the relativistic electron population.

\section{XBL TeV Source Candidates}

Within the SSC scenario justified above for Mrk 421,
we have used simple scaling arguments to predict the $\gamma$-ray fluxes in
different energy bands. 
A general property of the SSC mechanism is that the Compton component has
a spectrum which is similar to the synchrotron component, but upshifted
by $\sim\gamma_{e,max}'^2$ (up to the KN limit), 
where $\gamma_{e,max}'$ is the maximium electron Lorentz factor.
Thus, by comparing the synchrotron and Compton spectral components of Mrk 421,
which are both roughly parabolic on a logarithmic 
$\nu F_{\nu}$ plot (Macomb, \etal 1995),
we find an upshifting factor $\sim 10^9$ is required.
The implied value of $\gamma_{e,max} \sim 10^{4.5}$ is consistent with
that given in eq.(\ref{gamma}).
We note that the radio to optical and 0.1 to 1 GeV photon spectral indices
of the {\it EGRET} source XBLs are flatter than $E^{-2}$ (Vestrand, \etal 1995;
Sreekumar, \etal 1996) and the X-ray and Mrk 421 TeV 
spectra are steeper than $E^{-2}$ (Mohanty, \etal 1993; Petry, \etal 1996), as
expected for the parabolic spectral shapes.

We assume for simplicity that all XBLs have the same 
properties as those found for Mrk 421.
Both XBLs which have been detected by {\it EGRET}, Mrk421 and 
PKS2155-304, have $L_{\rm C}/L_{\rm syn}\sim 1$. We will assume that this 
ratio is the same for all XBLs.
The similarity between the synchrotron and Compton components,
with the upshifting factor of $\sim 10^9$ discussed
above, allows us to derive the following scaling law:
\begin{equation}
\label{scale}
\frac{\nu_oF_o}{L_{\rm syn}}\simeq\frac{\nu_{\rm GeV}F_{\rm GeV}}{L_{\rm C}}
\;\;{\rm and}\;\;
\frac{\nu_xF_x}{L_{\rm syn}}\simeq\frac{\nu_{\rm TeV}F_{\rm TeV}}{L_{\rm C}},
\end{equation}
From this equation, and assuming that $L_{\rm C}/L_{\rm syn}\sim 1$,
we obtain the energy fluxes for the GeV and TeV
ranges, 
\begin{equation}
\label{ef}
\nu_{\rm GeV}F_{\rm GeV} \sim \nu_oF_o \;\;{\rm and}\;\;
\nu_{\rm TeV}F_{\rm TeV} \sim \nu_xF_x
\end{equation}

In order to select good candidate TeV sources, we have used the {\it
EINSTEIN} slew survey sample given by Perlman, \etal (1996) to choose
low-redshift XBLs.
Using Eq.(\ref{ef}), we then calculated fluxes above 0.1
GeV for these sources.
We have normalized our calculations to the observed {\it EGRET} 
flux for Mrk 421. 
The energy fluxes $F_o$ and $F_x$ which we used in
the calculation are from Perlman, \etal (1996). 
The prime uncertainties in our calculations stem 
from our assumption that $(L_{C}/L_{syn}) \sim 1$ for all XBLs, from
the non-simultaneity of the data in different energy bands, and from the
fact that the synchrotron and Compton SEDs are not identical.
In order to calculate integral fluxes for these sources, we have assumed that
they have $E^{-1.8}$ photon spectra at energies between 0.1 and
10 GeV, the average spectral index for BL Lacs in this energy range.
We have also assumed an $E^{-2.2}$ photon source spectrum above 0.3 TeV
for all of these sources, based on preliminary data on Mrk 421 from 
the Whipple collaboration (Mohanty, \etal 1993).
We have taken account of
intergalactic absorption by using an optical depth which is an average 
between Models 1 and 2 of Stecker \& de Jager (1997).
Table 1 lists 23 XBLs at redshifts less than 0.2, giving our calculated
fluxes for these sources for energies above 0.1 GeV, 0.3 TeV and 1 TeV.

\section{Conclusions}

Within the context of a simple physical model, 
we have chosen 23 candidate TeV sources which are all nearby XBLs and have
predicted fluxes for these sources for energies above 0.1 GeV, 
0.3 TeV and 1 TeV.
Our calculations give fluxes which agree with all of the existing GeV and 
TeV \gray observations, including {\it EGRET} upper limits, to within
a factor of 2 to 3. 

Having normalized the Mrk 421 flux to a value of $1.43 \times 10^{-7}$
cm$^{-2}$s$^{-1}$ 
for $E_{\gamma} > 0.1$ GeV (Sreekumar, \etal 1996),
we predict a flux of $2.3\times10^{-11}$cm$^{-2}$s$^{-1}$ above 0.3 TeV.
This prediction is within 20\% of the average flux observed by the
Whipple collaboration over a four year time period (Schubnell, \etal 1996).
For Mrk 501, we predict a flux above 0.3 TeV which
should be observable with the Whipple telescope (as is indeed the case), 
whereas the corresponding 
0.1 GeV flux is predicted to be on the threshold of detection by {\it EGRET}.
(Mrk 501, as of this writing, has not been detected by {\it EGRET}.)
We predict a flux for PKS 2155-304 of $3.9\times10^{-7}$cm$^{-2}$s$^{-1}$ 
above 0.1 GeV.
For this source, a flux of $(2.7\pm0.7)\times10^{-7}$cm$^{-2}$s$^{-1}$ 
above 0.1 GeV
was detected during a single {\it EGRET} viewing period (Vestrand, \etal 1955),
close to our predicted value.
The tentative Whipple source 1ES2344+514 is one of our stronger source
predictions.
According to our calculations, PKS 2155-304, a southern hemisphere source 
which has not yet been looked at, should be relatively 
bright above 0.3 TeV, but not above 1 TeV, owing to intergalactic absorption.
Thus, TeV observations of this particular source may provide evidence for
the presence of intergalactic infrared radiation.


As Sambruna, \etal (1996) have pointed out, is is difficult to explain
the large differences in peak synchrotron frequencies between XBLs and
RBLs on the basis of jet orientation alone. The recent \gray evidence
discussed here suggests that similar large differences in peak Compton
energies
carry over into the \gray region of the spectrum via the SSC mechanism,
supporting the hypothesis that real physical differences exist between
XBLs (HBLs) and RBLs (LBLs).

\acknowledgments

We wish to acknowledge very helpful discussions with Carl Fichtel and Rita
Sambruna.





\begin{table}
\caption{Predicted $\gamma$-ray fluxes for low-redshift XBLs}
\begin{tabular}{|c|l|c|c|c|c|} 
\hline
1ES & Other &  z & $\phi (>0.1$ GeV) &
$\phi (>0.3$ TeV) & $\phi (>1$ TeV) \\
Name  & Name(s) &  & $10^{-7}$cm$^{-2}$s$^{-1}$ &
 $10^{-11}$cm$^{-2}$s$^{-1}$ &
$10^{-12}$cm$^{-2}$s$^{-1}$ \\
\hline
1ES$0145+138$   &       &  0.125  &  0.07 & 0.55 &  0.26 \\
1ES$0229+200$   &       &  0.139  &  0.08 & 0.28 &  0.11 \\
1ES$0323+022$   &  1H   &  0.147  &  0.11 & 0.40 &  0.15 \\
1ES$0347-121$   &       &  0.188 &  0.05 & 0.38 &  0.08 \\
1ES$0446+449$   &       &  0.203 &  0.04 & 0.09 &  0.02 \\
1ES$0548-322$   &PKS, 1H & 0.069 &  0.56 & 1.3  &  1.2  \\
1ES$0927+500$   &       &  0.188 &  0.06 & 0.12 &  0.02 \\
1ES$1101+384$   & Mrk 421 & 0.031 &  1.43 & 2.3  &  3.6  \\
1ES$1118+424$   & EXO     & 0.124 &  0.15 & 0.38 &  0.18 \\
1ES$1133+704$   &       & 0.046 &  1.5  & 0.94 &  1.2  \\
1ES$1212+078$   & Mrk 180, S5  & 0.136 &  0.22 & 0.07 &  0.03 \\
1ES$1239+069$   &         & 0.150 &  0.01 & 1.2  &  0.43 \\
1ES$1255+244$   &      &   0.141 &  0.41 & 0.88 &  0.34 \\
1ES$1312-423$   & MS     &  0.105 &  0.19 & 0.24 &  0.15 \\
1ES$1440+122$   &         &0.162 &  0.16 & 0.12 &  0.03 \\
1ES$1652+398$   & Mrk 501, S4  & 0.034 &  1.4  & 2.1  &  3.2  \\
1ES$1727+502$   & 1 Zw 187  &0.055 &  0.18 & 0.51 &  0.59 \\ 
1ES$1741+196$   &          & 0.083 &  0.21 & 0.43 &  0.35 \\
1ES$1959+650$   &          & 0.048 &  1.8  & 1.9  &  2.3  \\
1ES$2005-489$   & PKS      &0.071 &  0.70 & 0.91 &  0.84 \\
1ES$2155-304$   & PKS    &  0.116 &  3.9  & 1.7  &  0.88 \\
1ES$2321+419$   &        & 0.059 &  0.15 & 0.13 &  0.14 \\
1ES$2344+514$   &        & 0.044 &  0.54 & 0.61 &  0.80 \\
\hline      
\end{tabular}
\end{table}


\begin{thebibliography}{}
\bibitem[]{} Bloom, S.D.\& Marscher, A.P. 1993, in {\it AIP Conf. Proc.
280, Compton Gamma-Ray Observatory}, ed. M. Friedlander, N. Gehrels \&
D.J. Macomb (NewYork: AIP), p. 578.
\bibitem[]{} Celotti, A. \etal 1983, ApJ 416, 118.
\bibitem[]{} Dermer, C.D. \& Schlickeiser, R. 1993, ApJ 416, 458.
\bibitem[]{} Fichtel, C.E., \etal 1994, Ap.J.Suppl. 94, 551.
\bibitem[]{} Fichtel, C.E., \etal 1996, paper presented at Amer. Astro. Soc. meeting, Madison, WI, June, 1996.
\bibitem[]{} Ghisellini,G. \& Maraschi, L. 1989, ApJ 340, 181.
\bibitem[]{} Giommi, P. \& Padovani, P. 1994, MNRAS 268, L51.
\bibitem[]{} Kollgaard, R.I., Gabuzda, D.C. \& Feigelson, E.D. 1996, ApJ 460, 174.
\bibitem[]{} Lamer, G., Brunner, H. \& Staubert, R. 1996, submitted to A\&A.
\bibitem[]{} Macomb, D.J. \etal 1995, ApJ. 449, L99.
\bibitem[]{} Maraschi, L., Ghisellini, G., Tanzi, E.G. \& Treves, A. 1989,
ApJ 310, 325.
\bibitem[]{} Mattox, J.R., \etal 1996, submitted to Ap.J.
\bibitem[]{} Mohanty, G., \etal. 1993, {\it Proc. 23rd Intl Cosmic Ray Conf.}, University of Calgary Press, 1, 405.
\bibitem[]{} Padovani, P. \& Giommi, P. 1995, ApJ 444, 567.
\bibitem[]{} Padovani, P. \& Giommi, P. 1996, MNRAS 279, 526.
\bibitem[]{} Pearson, T.J. \& Zensus, J.A. 1987, in {\it Superluminal Radio Sources}, ed. Zensus, J.A. \& Pearson, T.J., Cambridge Univ. Press, p.1.
\bibitem[]{} Perlman, E.S., \etal 1996, Ap.J.Suppl. 104, 251.
\bibitem[]{} Petry, D. \etal 1996, A\&A, in press.
\bibitem[]{} Punch, M., \etal 1992, Nature 358, 477.
\bibitem[]{} Quinn, J., \etal 1995, ApJ 456, L83.
\bibitem[]{} Sambruna, R., Maraschi, L. and Urry, C.M. 1996, ApJ 463, 444.
\bibitem[]{} Sikora, M., Begelman, M.C. \& Rees, M.J. 1994, ApJ 421, 153.
\bibitem[]{} Sreekumar, P. \etal 1996, ApJ 464, 628.
\bibitem[]{} Stecker, F.W., De Jager, O.C. \& Salamon, M.H. 1992, ApJ 390, L49.
\bibitem[]{} Stecker, F.W. \& De Jager, O.C. 1997, Ap.J. 476, in press.
\bibitem[]{} Takahashi, T., \etal 1996, ApJ. Letters, in press.
\bibitem[]{} Thompson, \etal 1995, ApJ Suppl. 101, 259.
\bibitem[]{} Thompson, \etal 1996, ApJ Suppl., in press.
\bibitem[]{} Urry, C.M. \& Padovani, P. 1995, PASP 107,803.
\bibitem[]{} Vestrand, W.T., Stacy, J.G. \& Sreekumar, P. 1995, ApJ. 454, L96.
\end{thebibliography}
\end{document}